# Modeling and experimental study of the 27-day variation of galactic cosmic-ray intensity for a solar-wind velocity depending on heliolongitude

M.V. Alania [a,b], R. Modzelewska [a], A. Wawrzynczak [c],
[a] Institute of Math. And Physics of University of Podlasie, 3 Maja 54, 08-110 Siedlce, Poland
[b] Institute of Geophysics, Georgian Academy of Sciences, Tbilisi, Georgia
[c] Institute of Computer Science of University of Podlasie, 3 Maja 54, 08-110 Siedlce, Poland
(alania@ap.siedlce.pl, renatam@ap.siedlce.pl, awawrzynczak@ap.siedlce.pl).

*Abstract*

We develop a three dimensional (3-D) model of the 27-day variation of galactic cosmic ray (GCR) intensity with a spatial variation of the solar wind velocity. A consistent, divergence-free interplanetary magnetic field is derived by solving the corresponding Maxwell equations with a variable solar wind speed, which reproduces in situ observed experimental data for the time interval to be analyzed (24 August 2007-28 February 2008). We perform model calculations for the GCR intensity using the variable solar wind and the corresponding magnetic field. Results are compatible with experimental data; the correlation coefficient between our model predictions and observed 27-day GCR variation is $0.80 \pm 0.05$.

## 1. INTRODUCTION

Modeling of the 27-day variation of galactic cosmic ray (GCR) intensity is of interest since Richardson, Cane and Wibberenz (1999) found that the recurrent 27-day variation of solar wind parameters as well as that of GCR intensities are ~50% larger for positive (A>0) polarity epochs of solar magnetic cycles, than for the A<0. Previously, it was demonstrated (Alania et al., 2001a, 2001b; Gil and Alania 2001; Vernova et al., 2003; Iskra et al., 2004) that the amplitudes of the 27-day variation of the GCR intensity obtained from neutron monitors are greater in the minimum epochs of solar activity for the A>0 than for the A<0 epochs. Recently, we demonstrated (Alania et al., 2005, 2008a, 2008b; Gil et al., 2005) that also the amplitudes of the 27-day variation of the GCR anisotropy at solar minimum are greater when A>0 than when A<0.
Kota and Jokipii (2001) solved numerically a three-dimensional model of GCR transport and showed that larger recurrent variation of the proton flux can be expected in the positive A>0 periods than in the negative A<0 periods. Burger and Hitge (2004) and Burger et al., (2008) developed a hybrid model of the IMF, and suggested that the Fisk heliospheric magnetic field can explain several properties of the 27-day cosmic ray variation. Nevertheless, there are serious problems concerning the existence of the Fisk type heliospheric magnetic field (Fisk, 1996) in the minimum epochs of solar activity (Roberts et al., 2007).
It was demonstrated (Modzelewska et al., 2006; Alania et al., 2008a) that the heliolongitudinal distribution of the phase of the 27-day variation of the solar wind velocity has a clear maximum for the A>0 period, while it remains obscure for the A<0 period. The phase distribution of the 27–day variation of the solar wind velocity shows that a long–lived (~ 22 years) active heliolongitudes exist on the Sun preferentially for the A>0 polarity epoch of the solar magnetic cycle; the long–lived active heliolongitude is the source of the long-lived 27-day variation of the solar wind velocity, and afterwards, it can be considered as the general source of the 27-day variations of the GCR intensity and anisotropy. Moreover, Gil et al. (2008) showed that the amplitudes of the 27-day variation of the solar wind velocity are about two times greater for the A>0 epochs than for the A<0.

However, many of the papers (Gil and Alania, 2001; Alania et al., 2005; Gil et al., 2005; Alania et al., 2008a, 2008b; Kota and Jokipii, 2001; Burger and Hitge, 2004) aimed to explain results of Richardson et al. (1999), the general attention was paid to the drift effect and the role of recurrent changes of the solar wind velocity, which is a crucial (Alania et al., 2008a; Gil et al., 2008), was not considered.

To properly model the 27-day variation of the GCR intensity based on the Parker's (1965) transport equation the spatial and time dependences of the solar wind velocity $V$ and the interplanetary magnetic field (IMF) $B$ must be taken into account. However, it is rather complicated problem, because the validity of the Maxwell's equation $div B = 0$ should be kept for the time and spatially dependent solar wind velocity. Maxwell's equations for the IMF strength $B$ have a form (Parker, 1963):

$$\begin{cases} \dfrac{\partial B}{\partial t} = \nabla \times (V \times B) & (1a) \\ div B = 0 & (1b) \end{cases}$$

where $B$ is the IMF, $V$ –solar wind velocity, and $t$-time. The system of scalar equations for the components $(B_r, B_\theta, B_\varphi)$ of the IMF and components $(V_r, V_\theta, V_\varphi)$ of the solar wind velocity corresponding to Eqs. (1a) and (1b) can be rewritten in corotating frame (attached to the rotating Sun) in the heliocentric spherical $(r, \theta, \varphi)$ coordinate system, as:

$$\begin{cases} \dfrac{\partial B_r}{\partial t} = \dfrac{1}{r^2 \sin\theta}\left[\dfrac{\partial}{\partial \theta}[(V_r B_\theta - V_\theta B_r) r \sin\theta] - \dfrac{\partial}{\partial \varphi}[(V_\varphi B_r - V_r B_\varphi) r]\right] & (2a) \\ \dfrac{\partial B_\theta}{\partial t} = \dfrac{1}{r \sin\theta}\left[\dfrac{\partial}{\partial \varphi}(V_\theta B_\varphi - V_\varphi B_\theta) - \dfrac{\partial}{\partial r}[(V_r B_\theta - V_\theta B_r) r \sin\theta]\right] & (2b) \\ \dfrac{\partial B_\varphi}{\partial t} = \dfrac{1}{r}\left[\dfrac{\partial}{\partial r}[(V_\varphi B_r - V_r B_\varphi) r] - \dfrac{\partial}{\partial \theta}(V_\theta B_\varphi - V_\varphi B_\theta)\right] & (2c) \\ \dfrac{1}{r}\dfrac{\partial}{\partial r}(r^2 B_r) + \dfrac{1}{\sin\theta}\dfrac{\partial}{\partial \theta}(\sin\theta B_\theta) + \dfrac{1}{\sin\theta}\dfrac{\partial}{\partial \varphi} B_\varphi = 0 & (2d) \end{cases}$$

To solve Eqs (2a-2d) in general is difficult, but for our purpose, these equations can be simplified for the particular electro-magnetic conditions on the Sun and in the interplanetary space. Our aim in this paper is to compose a model of the 27-day variation of the GCR intensity for the changes of the solar wind velocity reproducing in situ measurements.

## 2. EXPERIMENTAL DATA

The simultaneous enhancements of the quasi periodic changes of the GCR intensity and parameters of solar wind were noticed by Richardson et al. (1999) for the positive polarity periods of the solar activity minima epochs. It was shown (Gil and Alania, 2001; Modzelewska et al., 2006; Alania et al., 2008a) that the heliolongitudinal asymmetry of the solar wind speed is one of the important sources of the 27-day variation of the GCR intensity and anisotropy.

In this paper we analyze experimental data of the daily solar wind velocity, GCR intensity from the Moscow neutron monitor and radial $B_x$, azimuthal $B_y$ and heliolatitudinal $B_z$ components of the IMF for the period of 24 August 2007-28 February 2008 (Fig 1).

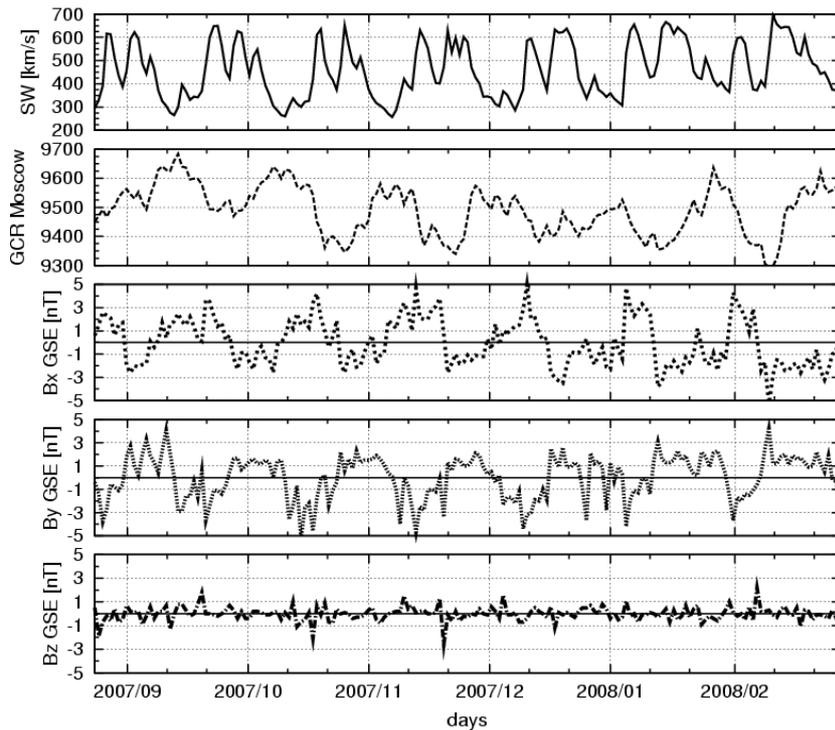

Fig. 1 Temporal changes of the daily solar wind velocity (SW) [OMNI], GCR intensity by Moscow neutron monitor and radial $B_x$, azimuthal $B_y$ and latitudinal $B_z$ components of the IMF [OMNI] for the period of 24 August 2007-28 February 2008

Fig. 1 shows that the quasi periodic changes are clearly established in all parameters except for the $B_z$ component of the IMF. The solar wind velocity V is in opposite correlation with the changes of the GCR intensity. There is not any recognizable relation of the changes of the $B_z$ component (due to its negligible values) with other parameters; also, it is obvious that the contribution of the $B_z$ component in the changes of the magnitude of the IMF is negligible. At the same time the solar wind velocity V undoubtedly shows an existence of the first (27 days), second (14 days) and third (9 days) harmonics. Generally higher harmonics in the solar rotation period (e.g. 14 and 9 days) are related with the simultaneous existence of several active heliolongitudes (Alania and Shatashvili, 1974). Recently Temmer et al., (2007) provided evidence that the 9-day period in the solar wind parameters might be caused by the periodic longitudinal distribution of coronal holes on the Sun recurring for several Carrington rotations. As it was shown (Gil and Alania, 2001; Modzelewska et al., 2006; Alania et al., 2008a), the heliolongitudinal asymmetry of the solar wind velocity is one of the crucial parameters in creation of the 27-day variation of the GCR intensity. In connection with this, we estimate the contribution of each of the three harmonics (27, 14 and 9 days) in the daily changes of the solar wind velocity. We use the frequency filter method (e.g., Otnes and Enochson, 1972). This technique decomposes a time series into frequency components. We use band pass filter characterized by two period (frequency) bounds which transmits only the components with a period (frequency) within these bounds. A band-pass filter rejects high and low frequencies, passing only signal around some intermediate frequency. The frequency-domain behavior of a filter is described mathematically in terms of its transfer function or network function. This is the ratio of the Laplace transforms of its output and input signals. We investigate periodicity bound within 24-32 days (27-28 days in the middle) for the I harmonic, 11-17 days (14 days in the middle) for II harmonic and 6-12 days (9 days in the middle) for III harmonic of the 27-day wave. For comparison, this procedure have been performed for all parameters - GCR intensity by Moscow neutron monitor, radial $B_x$, azimuthal $B_y$ and latitudinal $B_z$ components of the IMF for the period of 24 August 2007-28 February 2008 (Figs. 2abcde).

Presented in Figs 2a-2e are the temporal changes of all considered parameters (Fig.2a corresponds to the solar wind velocity, Fig. 2b - the GCR intensity, Fig. 2c, Fig. 2d, and Fig. 2e - to the $B_x$, $B_y$ and $B_z$ components of the IMF, respectively) and the first (27 days) harmonic wave (upper panels), the sum of the first (27 days) and the second (14 days) harmonic waves (middle panels), and the sum of the first (27 days), second (14 days), and third (9 days) harmonic waves (bottom panels).

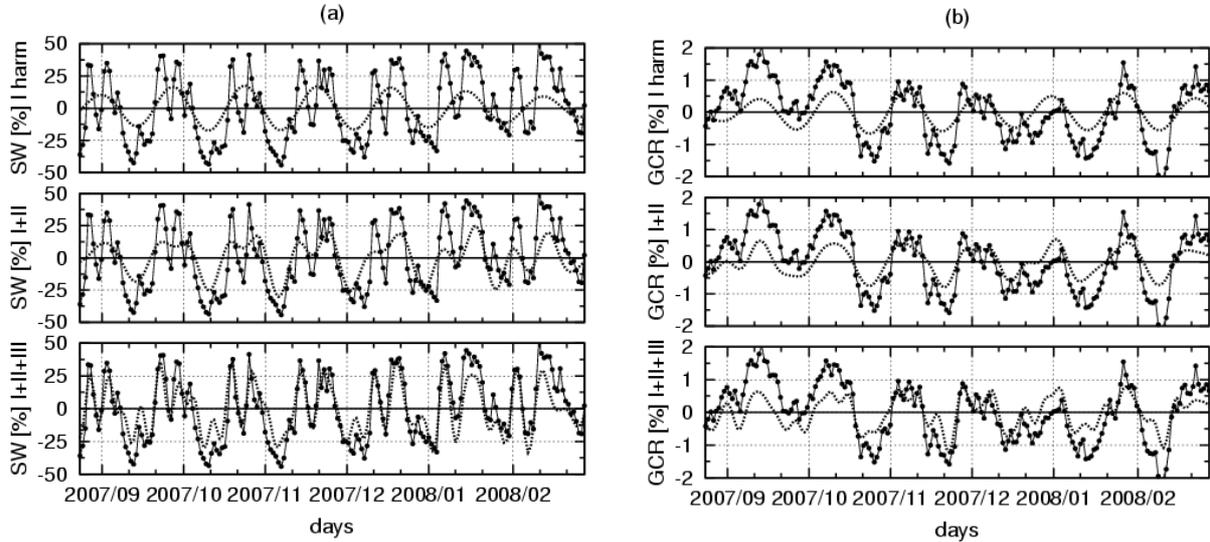

Fig. 2a. In each (top, middle and bottom) panel are presented temporal changes of the daily solar wind velocity (solid lines) and changes of the first harmonic wave (top panel), the sum of I and II harmonic waves (middle panel), and the sum of I, II and III harmonic waves (bottom panel) for the period of 24 August 2007-28 February 2008.

Fig. 2b. As in Fig. 2a but for the GCR intensity measured by Moscow neutron monitor

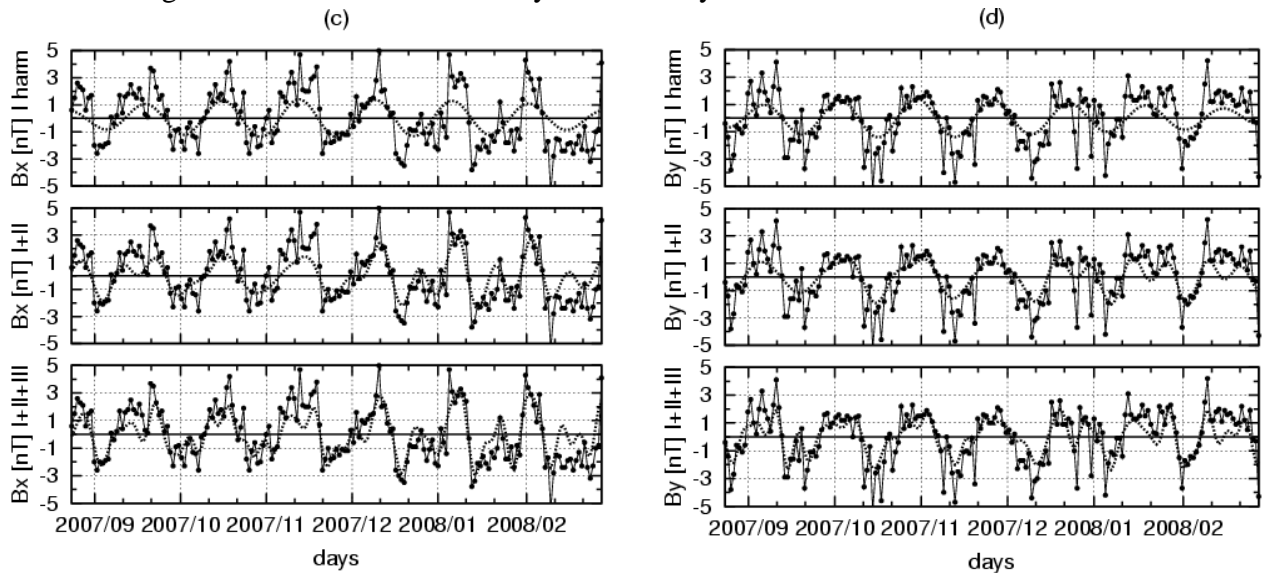

Fig. 2c. As in Fig. 2a but for the radial $B_x$ component of the IMF.
Fig. 2d. As in Fig. 2a but for the azimuthal $B_y$ component of the IMF.

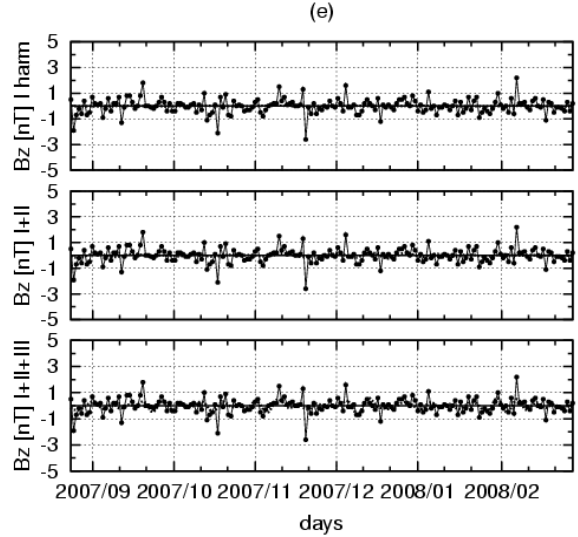

Fig. 2e. As in Fig. 2a but for the latitudinal component $B_z$ component of the IMF.

To estimate quantitatively the contributions of each harmonic (27, 14 and 9 days) in parameters presented in Fig. 2abcde, we calculated the correlation coefficients between the observed data and first harmonic waves, the observed data and the sum of I and II harmonic waves, the observed data and the sum of I, II and III harmonic waves for the period of 24 August 2007-28 February 2008. Results of calculations are presented in Table 1.

Table 1

|                   | SW        | GCR       | Bx_GSE    | By_GSE    | Bz_GSE    |
|-------------------|-----------|-----------|-----------|-----------|-----------|
| I harm wave       | 0.69±0.05 | 0.75±0.05 | 0.69±0.05 | 0.71±0.05 | 0.22±0.07 |
| I+II harm wave    | 0.74±0.05 | 0.77±0.05 | 0.80±0.04 | 0.76±0.05 | 0.28±0.07 |
| I+II+III harm wave| 0.87±0.04 | 0.77±0.05 | 0.82±0.04 | 0.79±0.05 | 0.46±0.06 |

We see from the Table 1:
a) the 27-day periodicity dominates in the changes of the solar wind velocity. The correlation coefficient between observed data of the solar wind velocity and its first harmonic wave equals 0.69±0.05, an inclusion of the second harmonic (14 days) slightly increases the correlation coefficient (0.74±0.05), and the contribution of the third harmonic (9 days) is also significant; the correlation coefficient increases up to 0.87±0.04;
b) the 27-day periodicity dominates in the changes of the GCR intensity. The correlation coefficient between observed data of the GCR intensity and its first harmonic wave equals 0.75±0.05, while an inclusion of the second (14 days) and the third (9 days) harmonics in reality do not contribute at all; the correlation coefficients equal 0.77±0.05;
c) the 27-day periodicity dominates in the changes of $B_x$ component of the IMF. The correlation coefficient between observed data of $B_x$ and its first harmonic wave (27 days) equals 0.69±0.05; after inclusion of the second harmonic (14 days) the correlation coefficient increases up to 0.80±0.04, but the contribution of the third harmonic (9 days) is negligible; the correlation coefficient equals 0.82±0.04;
d) the 27-day periodicity dominates in the changes of $B_y$ component of the IMF. The correlation coefficient between observed data of $B_y$ and its first harmonic wave (27 days) equals 0.71±0.05; an inclusion of the second harmonic (14 days) slightly increases the correlation coefficient (0.76±0.05), but the contribution of the third harmonic (9 days) is not noticeable; the correlation coefficient equals 0.79±0.05;
e) the 27-day periodicity is almost absent in the changes of $B_z$ component of the IMF. The correlation coefficient between observed data of $B_z$ and its first harmonic wave (27 days) equals 0.22±0.07; there is some contribution of the second harmonic (correlation coefficient is equal to 0.28±0.07), but the contribution of the third harmonic is more valuable; the correlation coefficient increases up to 0.46±0.06.

The temporal changes of each parameter are similar (quasi steady) from one Carrington rotation to another. So, the changes of the solar wind velocity, the GCR intensity, $B_x$, $B_y$, $B_z$ components of the IMF with the solar rotation period can be considered as a quasi stationary during seven Carrington rotations period (24 August 2007-28 February 2008). The averaged values of all parameters by means of seven Carrington rotations daily data are presented in Fig. 3.

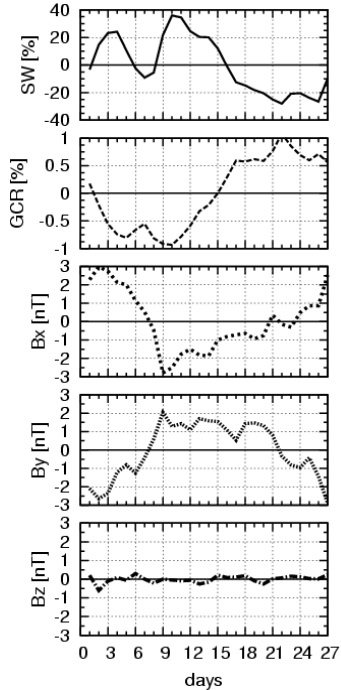

Fig. 3. Temporal changes of the daily data of the SW velocity, GCR intensity measured by Moscow neutron monitor, $B_x$, $B_y$, $B_z$ components of the IMF in GSE system, averaged during 7 Carrington rotations for the period of 24 August 2007-28 February 2008.

Fig.3 shows that the changes of the $B_z$ component of the IMF is negligible (its value oscillates near zero) in comparison with the changes of the $B_x$ and $B_y$ components. Correlation between the changes of the solar wind velocity and the GCR intensity during Carrington rotation (Fig. 3) is negative (-0.84±0.05).
This high anticorrelation shows the importance of the heliolongitudinal dependence of the solar wind velocity in causing the 27-day variation of the GCR intensity. Among any other pairs of parameters the direct correlations are insignificant.

## 3 MODEL OF THE 27-DAY VARIATION OF THE GCR INTENSITY

*3.1 Numerical solution of Maxwell's equation*

The changes of the solar wind velocity, the GCR intensity, $B_x$, $B_y$ and $B_z$ components of the IMF are quasi stationary for seven Carrington rotations during the time interval to be analyzed (24 August 2007-28 February 2008). So, we assume that $\frac{\partial B_r}{\partial t} = 0$, $\frac{\partial B_\theta}{\partial t} = 0$, $\frac{\partial B_\varphi}{\partial t} = 0$ in Eqs. (2a)-(2c). We also accept that average value of the heliolatitudinal component of the solar wind velocity $V_\theta$ equals zero; then the system of Eqs. (2a)-(2d) can be reduced, as

$$\begin{cases} \sin\theta\, V_r \dfrac{\partial B_\theta}{\partial \theta} + \sin\theta\, B_\theta \dfrac{\partial V_r}{\partial \theta} + \cos\theta\, V_r B_\theta - V_\varphi \dfrac{\partial B_r}{\partial \varphi} - B_r \dfrac{\partial V_\varphi}{\partial \varphi} + V_r \dfrac{\partial B_\varphi}{\partial \varphi} + B_\varphi \dfrac{\partial V_r}{\partial \varphi} = 0 & (3a) \\ V_\varphi \dfrac{\partial B_\theta}{\partial \varphi} + B_\theta \dfrac{\partial V_\varphi}{\partial \varphi} + r\sin\theta\, V_r \dfrac{\partial B_\theta}{\partial r} + r\sin\theta\, B_\theta \dfrac{\partial V_r}{\partial r} + \sin\theta\, V_r B_\theta = 0 & (3b) \\ rB_r \dfrac{\partial V_\varphi}{\partial r} + rV_\varphi \dfrac{\partial B_r}{\partial r} + V_\varphi B_r - V_r B_\varphi - rV_r \dfrac{\partial B_\varphi}{\partial r} - rB_\varphi \dfrac{\partial V_r}{\partial r} + B_\theta \dfrac{\partial V_\varphi}{\partial \theta} + V_\varphi \dfrac{\partial B_\theta}{\partial \theta} = 0 & (3c) \\ \dfrac{\partial B_r}{\partial r} + \dfrac{2}{r} B_r + \dfrac{ctg\,\theta}{r} B_\theta + \dfrac{1}{r} \dfrac{\partial B_\theta}{\partial \theta} + \dfrac{1}{r\sin\theta} \dfrac{\partial B_\varphi}{\partial \varphi} = 0 & (3d) \end{cases}$$

The latitudinal component $B_\theta$ of the IMF is very weak for the period to be analyzed, so we can assume that it equals zero $(B_\theta = 0)$. This assumption straightforwardly leads (from Eq. (2a)) to the relationship between $B_r$ and $B_\varphi$, as, $B_\varphi = -B_r \dfrac{V_\varphi}{V_r}$, where $V_\varphi = \Omega r \sin\theta$ is the corotational speed. Then Eq. (3d) with respect to the radial component $B_r$ has a form:

$$A_1 \frac{\partial B_r}{\partial r} + A_2 \frac{\partial B_r}{\partial \varphi} + A_3 B_r = 0 \qquad (4)$$

The coefficients $A_1$, $A_2$ and $A_3$ depend on the radial $V_r$ and heliolongitudinal $V_\varphi$ components of the solar wind velocity V.

Our goal is to solve Eq. (4) in heliocentric coordinate system $(r, \theta, \varphi)$ for the changeable solar wind velocity reproducing in situ measurements in the interplanetary space. We demonstrated (Fig. 2a) that the sum of three harmonics (27, 14, 9 days) sufficiently describes the temporal changes of the solar wind velocity (correlation coefficient between observed data and sum of first, second and third harmonic waves equals 0.87±0.04, see Table 1). Presented in Fig.4 are the averaged values of the solar wind velocity (points) for seven Carrington rotations daily data (points) and dashed curve representing the sum of three harmonic waves.

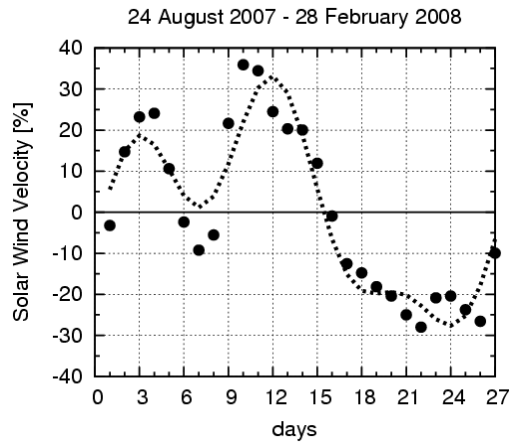

Fig. 4 Temporal changes of the solar wind velocity (points) by means of seven Carrington rotations daily data (points) and dashed curve representing the sum of three harmonic (27, 14 and 9 days) waves for the period of 24 August 2007-28 February 2008.

We include in Eq. (4) approximation of the changes of the average solar wind speed calculated from the experimental data (Fig. 4):

$$V_r = 400 \times (1 + \alpha_1 \sin(\varphi - 0.5) + \alpha_2 \sin(2(\varphi - 0.9)) + \alpha_3 \sin(3(\varphi - 0.15))) \quad (5)$$

where, $\varphi$ is heliolongitude and $\alpha_1 = 0.2275$ $\alpha_2 = -0.0925$ $\alpha_3 = 0.106$

We take into account, as well:

$$V_\theta = 0, \quad V_\varphi = \Omega r \sin\theta \quad (6)$$

Taking into consideration the expressions (5) and (6) the coefficients $A_1$, $A_2$ and $A_3$ in Eq. (4) are

$$A_1 = 1$$

$$A_2 = -\frac{\Omega}{400 \times (1 + \alpha_1 \sin(\varphi - 0.5) + \alpha_2 \sin(2(\varphi - 0.9)) + \alpha_3 \sin(3(\varphi - 0.15)))}$$

$$A_3 = \frac{2}{r} + \frac{400 \times \Omega \times (\alpha_1 \cos(\varphi - 0.5) + 2\alpha_2 \cos(2(\varphi - 0.9)) + 3\alpha_3 \cos(3(\varphi - 0.15)))}{(400 \times (1 + \alpha_1 \sin(\varphi - 0.5) + \alpha_2 \sin(2(\varphi - 0.9)) + \alpha_3 \sin(3(\varphi - 0.15))))^2}$$

Equation (4) is first order linear partial differential equation. It can be solved analytically (e.g. Polyanin et al., 2002), as well by numerical method. We solve Eq. (4) by numerical method.

Equation (4) was reduced to the algebraic system of equations using a difference scheme method (e.g., Kincaid and Cheney, 2006), as

$$A_1 \frac{B_r[i+1,j,k] - B_r[i,j,k]}{\Delta r} + A_2 \frac{B_r[i,j,k+1] - B_r[i,j,k]}{\Delta \varphi} + A_3 B_r[i,j,k] = 0 \quad (7)$$

where, i=1,2,…, I; j=1,2,…, J; k=1,2,…, K are steps in radial distance, vs. heliolatitude and heliolongitude, respectively. Then Eq. (7) was solved by the iteration method with the boundary condition near the Sun $B_r[1,j,k] = const$; in considered case $r_1 = 0.5$ AU, $B_r[1,j,k] = \begin{cases} 18 \ nT & for \quad 0^0 \leq \theta \leq 90^0 \\ -18 \ nT & for \quad 90^0 < \theta \leq 180^0 \end{cases}$ for the positive polarity period (A>0).

The choice of these boundary conditions was stipulated by requiring agreement of the solutions of Eq. (7) with the in situ measurements of the $B_r$ and $B_\varphi$ components of the IMF at the Earth orbit.

Results of the solution of Eq. (7) for the $B_r$ and $B_\varphi$ components of the IMF calculated by the expression $B_\varphi = -B_r \frac{V_\varphi}{V_r}$ are presented in Figs. 5-7, respectively.

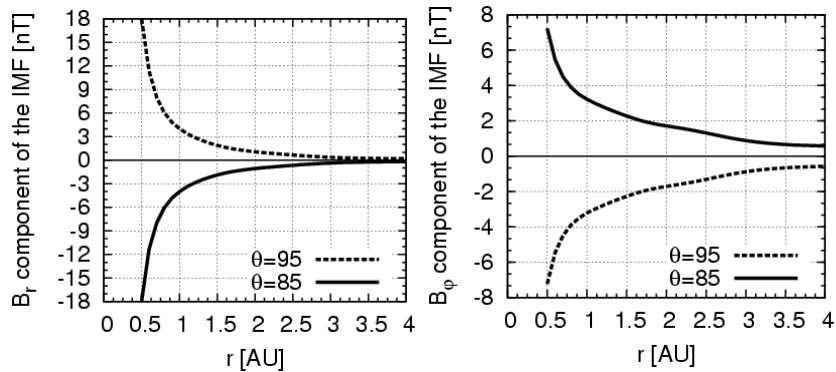

Fig. 5 Radial changes of the $B_r$ and $B_\varphi$ components of the IMF for different heliolatitudes near the solar equatorial plane

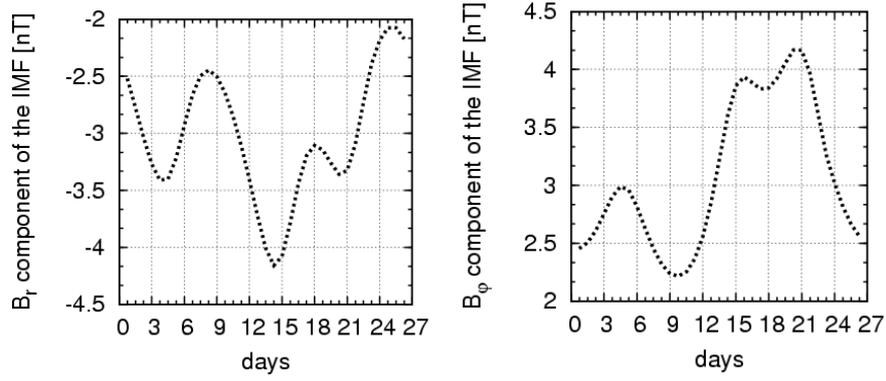

Fig. 6 Azimuthal changes of the $B_r$ and $B_\varphi$ components of the IMF at the Earth orbit.

Presented in Fig. 7 are the heliolatitudinal variation of the $B_\varphi$ component of the IMF at 1 AU, calculated by the formula $B_\varphi = -B_r \dfrac{V_\varphi}{V_r}$.

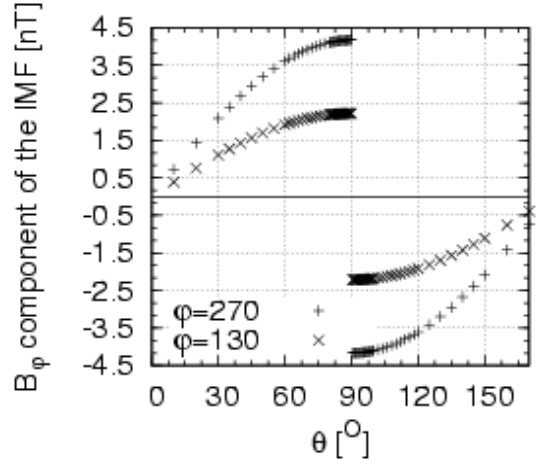

Fig. 7 Heliolatitudinal changes of the $B_\varphi$ component of the IMF at 1 AU for $\varphi = 130°$ and $\varphi = 270°$.

*3.2 Modeling of the 27-day variation of the GCR intensity.*

For modeling the 27-day variation of the GCR intensity we use stationary $\left(\dfrac{\partial N}{\partial t} = 0\right)$ Parker's transport equation (Parker, 1965):

$$\nabla_i (K_{ij} \nabla_j N) - \nabla_i (V_i N) + \frac{1}{3} \frac{\partial}{\partial R}(NR) \nabla_i V_i = 0 \qquad (8)$$

Where $N$ and $R$ are density and rigidity of cosmic ray particles, respectively; $V_i$ – solar wind velocity, $K_{ij}$ is the anisotropic diffusion tensor of galactic cosmic rays. The parallel diffusion coefficient $\kappa_\parallel$ changes versus the spatial spherical coordinates ($r, \theta, \varphi$) and rigidity R of GCR particles as, $\kappa_\parallel = \kappa_0 \kappa(r) \kappa(R)$, where $\kappa_0 = \frac{\lambda_0 v}{3} = 2 \times 10^{22}$ cm²/s, $\kappa(r) = 1 + \alpha_0 r$, $\alpha_0 = 0.5$, $v$ is the velocity of GCR particles, $\kappa(R) = (R/1GV)^{0.5}$. So, the parallel diffusion coefficient for the GCR particles of 10 GV rigidity equals, $\kappa_\parallel \approx 10^{23}$ cm²/s at the Earth orbit. The ratios of $\beta$ and $\beta_1$ of the perpendicular and drift diffusion coefficients

$\kappa_\perp$ and $\kappa_d$ to the parallel diffusion coefficient $\kappa_\parallel$ are in standard form: $\beta = \frac{\kappa_\perp}{\kappa_\parallel} = (1+\omega^2\tau^2)^{-1}$ and $\beta_1 = \frac{\kappa_d}{\kappa_\parallel} = \omega\tau(1+\omega^2\tau^2)^{-1}$, where $\omega\tau = 300\,B\,\lambda_0\,R^{-1}$, B is the strength of the IMF and $\lambda_0$ - the transport free path of GCR particles and for the particles with rigidity 10 GV at the Earth orbit $\beta \approx 0.1$ and $\beta_1 \approx 0.3$.

In this model we assume that the stationary 27-day variation of the GCR intensity is caused by the heliolongitudinal asymmetry of the solar wind speed. In Eq. (8), we included the $B_r$ and $B_\varphi$ components of the IMF obtained from the numerical solution of Eq. (7), and the changes of the solar wind velocity (5), as well. Implementation of the heliospheric magnetic field obtained from the numerical solution of Eq. (7) in Parker's transport equation is done through the diffusion coefficients and spiral angle $\psi = \arctan\left(-\frac{B_\varphi}{B_r}\right)$ in anisotropic diffusion tensor of GCR particles ($\psi$ is the angle between magnetic field lines and radial direction in the equatorial plane).

The kinematical model of solar wind has some limitations; in particular it can be applied until some radial distance where the faster wind would overtake the previously emitted slower one. To avoid an intersection of the IMF lines the heliolongitudinal asymmetry of the SW velocity takes place only up to the distance of ~ 8 AU and then $V = 400\,km/s$ with the Parker's spiral of the IMF throughout the heliosphere. Equation (8) was solved numerically as in our papers published elsewhere (e.g., Alania, 2002; Iskra et al., 2004; Modzelewska et al., 2006). Changes of the relative density obtained as a solution of the transport Eq. (8) for the model of the 27-day variation are presented in Fig. 8 (dashed line); in this figure are also presented (points) changes of the GCR intensity obtained by Moscow neutron monitor experimental data averaged for 7 Carrington rotations during the period of 24 August 2007-28 February 2008 (Fig. 1), as well.

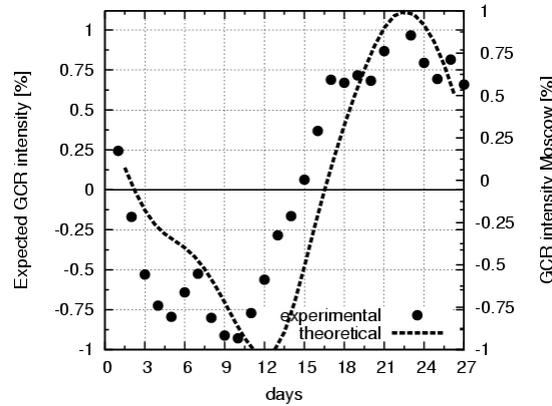

Fig. 8 Heliolongitudinal changes of the expected GCR intensity for rigidity 10GV at the Earth orbit during solar rotation period (dashed line) and temporal changes of superimposed GCR intensity by Moscow neutron monitor during 27 day for the period of 24 August 2007-28 February 2008 (points).

Fig. 8 shows that results of theoretical modeling (dashed line) and the experimental data (points) are in good agreement; the correlation coefficient between the 27-day variations of the GCR intensity observed (averaged for seven Carrington rotations for the period of 24 August 2007-28 February 2008) by Moscow neutron monitor and expected from the proposed model of the 27-day variation of the GCR intensity equals 0.80 ± 0.05. We underline that the presented model of the 27-day variation of the GCR intensity composed for the changeable solar wind velocity (5) and the components $B_r$ and $B_\varphi$ obtained as the solutions of Eq. (7) is compatible with the experimental data.

## 4. CONCLUSIONS

1. The quasi steady 27-day variations of the solar wind velocity, GCR intensity, and $B_r$ and $B_\varphi$ components of the IMF have been analyzed for seven succession Carrington rotations in the period of 24 August 2007-28 February 2008.
2. The Maxwell equations are solved with a solar wind speed varying in heliolongitude in accordance with in situ measurements to derive the longitudinal dependence of the $B_r$ and $B_\varphi$ components of the IMF.
3. A three-dimensional model is proposed for the 27-day variation of GCR intensity in response to a realistic variation of the solar wind velocity. The model incorporates the $B_r$ and $B_\varphi$ components of the IMF derived from solving the Maxwell equations.
4. The proposed model of the 27-day variation of the GCR intensity is in good agreement with the observational material. The correlation coefficient between the 27-day variation of the GCR intensity observed (averaged for seven Carrington rotations for the period of 24 August 2007-28 February 2008) by Moscow neutron monitor and the predictions from our model is $0.80 \pm 0.05$.

ACKNOWLEDGMENTS

Authors thank the investigators of Moscow neutron monitor station, SPDF OMNIWeb database for possibility to use their data.
Authors kindly benefited from discussions with Prof. R. A. Burger at COSPAR meeting in Montreal.
Renata Modzelewska would like to thank COSPAR Committee for partial sponsorship to attend 37[th] COSPAR Scientific Assembly 2008.
This work has been supported by Foundation for Polish Science.
Author thanks for the referee remarks which were very helpful.

REFERENCES

Alania, M. V., Gil, A., Modzelewska, R., Study of the 27-day variations of the galactic cosmic ray intensity and anisotropy, Adv. Space Res. 41, 1, 280 – 286, 2008a.

Alania, M.V., Gil, A., Modzelewska, R., 27-day variations of the galactic cosmic ray intensity and anisotropy, Astrophysics and Space Sciences Transactions, 4, 1, 31-34, 2008b.

Alania, M. V., Gil, A., Iskra, K., Modzelewska, R., 27-day variations of the galactic cosmic ray intensity and anisotropy in different solar magnetic cycles (1964-2004), Proc. 29th ICRC, Pune, 2, 215-218, 2005.

Alania, M. V., Stochastic Variations of Galactic Cosmic Rays, Acta Phys. Polonica B, 33, 4, 1149-1166, 2002

Alania, M. V., Vernova, E. S., Tyasto, M. I., Baranov, D. G., Experimental and modeling investigation of 27-day variation of galactic cosmic rays, Izvestia RAN, 65, 3, 370 – 372, 2001a

Alania M. V., D. G. Baranov, M. I. Tyasto and E. S. Vernova, 27-Day variations of galactic cosmic rays and changes of solar and geomagnetic activities, Adv. Space Res., 27, 3, 619-624, 2001b

Alania, M. V., Shatashvili, L. Kh., Quasi-periodic Cosmic Ray variations, Mecniereba, Tbilisi, 1974 (In Russian)

Burger, R.A., Kruger, T. P. J., Hitge, M., Engelbrecht, N. E., A Fisk-Parker Hybrid Heliospheric Magnetic Field With a Solar-Cycle Dependence, Astrophys. J., 674, 511-519, 2008

Burger, R.A., Hitge, M. The Effect of a Fisk-Type Heliospheric Magnetic Field on Cosmic-Ray Modulation, Astrophys. J., 617, L73-L76, 2004

Fisk, L. A., Motion of the footpoints of heliospheric magnetic field lines at the Sun: Implications for recurrent energetic particle events at high heliographic latitudes, J. Geophys. Res., 101, A7, 15547-15554, 1996

Gil, A., Alania, M.V., Modzelewska, R., On the quasi-periodic variations of the galactic cosmic rays intensity and anisotropy in the lingering minimum of solar activity, Proc. 21$^{st}$ ECRS, Kosice, Slovakia, 2008

Gil, A., Alania, M.V., Modzelewska, R., Iskra, K., On the 27-day variations of the galactic cosmic ray anisotropy and intensity for different periods of solar magnetic cycle, Adv. Space Res., 35, 4, 687–690, 2005.

Gil, A., Alania, M. V., 27-day variation of cosmic rays for the minima epochs of solar activity: experimental and 3-D drift modeling results, Proc. 27th ICRC, Hamburg, 9, 3725-3728, 2001.

Iskra K., Alania, M. V., Gil, A., Modzelewska, R., Siluszyk, M., On the roles of the stochastic and regular heliospheric magnetic fields in different classes of galactic cosmic rays variations, Acta Phys. Polonica B, 35, 4, 1565-1580, 2004.

Kincaid, D., Cheney, W., "Numerical Analysis", WNT, Warsaw, 2006 (in Polish)

Kota, J., Jokipii, J.R., Recurrent depressions of galactic cosmic rays in CIRs: 22-year cycle, Proc. 27th Inter. Cosmic Ray Conf., 9, 3577- 3580, Hamburg, 2001

Modzelewska, R., Alania, M.V., Gil, A., Iskra, K., 27-day variations of the galactic cosmic ray intensity and anisotropy, Acta Phys. Pol. B, 37, 5, 1641-1650, 2006

Otnes, R. K., Enochson, L., Digital Time Series Analysis, John Wiley and Sons, New York, 1972

Parker, E. N., The passage of energetic charged particles through interplanetary space, Planet. and Space Sci., 13, 9-49, 1965.

Parker, E. N., "Interplanetary Dynamical Processes", Interscience Publishers, New York, 1963.

Polyanin, D., Zaitsev, V. F., Moussiaux, A., Handbook of First Order Partial Differential Equations, Taylor & Francis, London, 2002.

Richardson, I. G., Cane, H. V., Wibberenz, G. A., A 22-year dependence in the size of near-ecliptic corotating cosmic ray depressions during five solar minima, J. Geophys. Res. 104, 12549–12553, 1999.

Roberts, D. A., Giacalone, J., Jokipii, J. R.; Goldstein, M. L.; Zepp, T. D., Spectra of polar heliospheric fields and implications for field structure, J. Geophys. Res., 112, A8, CiteID A08103, doi:10.1029/2007JA012247, 2007

Temmer, M., Vršnak, B., Veronig, A. M., Periodic Appearance of Coronal Holes and the Related Variation of Solar Wind Parameters, Solar Phys., 241, 2, 371-383, 2007.

Vernova, E. S., Tyasto, M. I., Baranov, D. G., Alania, M. V., Gil, A., On the influence of solar activity and the solar magnetic field on the 27-day variation of galactic cosmic rays, Adv. Space Res., 32, 4, 621-626, 2003.